\documentclass{desyproc}

\begin{document}
\title{The Rethermalizing Bose-Einstein Condensate of 
Dark Matter Axions}

\author{{\slshape  Nilanjan Banik, Adam Christopherson, 
Pierre Sikivie, Elisa Maria Todarello}\\[1ex]
University of Florida, Gainesville, FL 32611, USA}

\contribID{familyname\_firstname}

\confID{11832}  
\desyproc{DESY-PROC-2015-02}
\acronym{Patras 2015} 
\doi  

\maketitle

\begin{abstract}
The axions produced during the QCD phase transition by vacuum 
realignment, string decay and domain wall decay thermalize as 
a result of their gravitational self-interactions when the 
photon temperature is approximately 500 eV.  They then form a 
Bose-Einstein condensate (BEC).  Because the axion BEC 
rethermalizes on time scales shorter than the age of the 
universe, it has properties that distinguish it from other 
forms of cold dark matter.  The observational evidence for 
caustic rings of dark matter in galactic halos is explained 
if the dark matter is axions, at least in part, but not if 
the dark matter is entirely WIMPs or sterile neutrinos.
\end{abstract}

\section{Axion dark matter}

The story we tell applies to any scalar or pseudo-scalar dark 
matter produced in the early universe by vacuum realignment
and/or the related processes of string and domain wall decay.  
However, the best motivated particle with those properties  
is the QCD axion since it is not only a cold dark matter 
candidate but also solves the strong CP problem of the 
standard model of elementary particles \cite{PQ,WW}.  So,
for the sake of definiteness, we discuss the specific case
of the QCD axion. 

The Lagrangian density for the axion field $\phi(x)$ may be written 
as 
\begin{equation}
{\cal L}_a = {1 \over 2} \partial_\mu \phi \partial^\mu \phi 
- {1 \over 2} m^2 \phi^2 + {\lambda \over 4!} \phi^4 + ...
\label{lagr}
\end{equation}
where the dots represent interactions of the axion with the 
known particles.  The properties of the axion are mainly determined 
by one parameter $f$ with dimension of energy, called the `axion 
decay constant'.  In particular the axion mass is
\begin{equation}
m \simeq {f_\pi m_\pi \over f} {\sqrt{m_u m_d} \over m_u + m_d}
\simeq 6 \cdot 10^{-6} {\rm eV}~{10^{12}~{\rm GeV} \over f}
\label{axm}
\end{equation} 
in terms of the pion decay constant $f_\pi$, the pion mass $m_\pi$
and the masses $m_u$ and $m_d$ of the up and down quarks, and the 
axion self-coupling is
\begin{equation}
\lambda \simeq {m^2 \over f^2} 
{m_d^3 + m_u^3 \over (m_u + m_d)^3} \simeq
0.35 {m^2 \over f^2}~~\ .
\label{axl}
\end{equation}
All couplings  of the axion are inversely proportional to $f$.  When 
the axion was first proposed, $f$ was thought to be of order the 
electroweak scale, but its value is in fact arbitrary \cite{ZZ}.  
However the combined limits from unsuccessful searches for the 
axion in particle and nuclear physics experiments and from stellar 
evolution imply $f \gtrsim 3 \cdot 10^9$ GeV \cite{axrev}.

An upper limit $f \lesssim 10^{12}$ GeV is obtained from the requirement 
that axions are not overproduced in the early universe by the vacuum 
realignment mechanism \cite{axdm}, which may be briefly described as 
follows.  The non-perturbative QCD effects that give the axion its mass 
turn on at a temperature of order 1 GeV.  The critical time, defined by 
$m(t_1) t_1 = 1$, is 
$t_1 \simeq 2 \cdot 10^{-7}~{\rm sec}(f/10^{12}~{\rm GeV})^{1 \over 3}$.  
Before $t_1$, the axion field $\phi$ has magnitude of order $f$.  After 
$t_1$, $\phi$ oscillates with decreasing amplitude, consistent with 
axion number conservation.  The number density of axions 
produced by vacuum realignment is  
\begin{equation}
n(t) \sim {f^2 \over t_1}\left({a(t_1) \over a(t)}\right)^3 = 
{4 \cdot 10^{47} \over {\rm cm}^3}
\left({f \over 10^{12}~{\rm GeV}}\right)^{5 \over 3}
\left({a(t_1) \over a(t)}\right)^3
\label{axden}
\end{equation}
where $a(t)$ is the cosmological scale factor.  Their contribution to 
the energy density today equals the observed density of cold dark matter 
when the axion mass is of order $10^{-5}$ eV, with large uncertainties.
The axions produced by vacuum realignment are a form of cold dark matter
because they are non-relativistic soon after their production at time $t_1$.  
Indeed their typical momenta at time $t_1$ are of order $1/t_1$, and vary 
as $1/a(t)$, so that their velocity dispersion is
\begin{equation}
\delta v(t) \sim {1 \over m t_1} {a(t_1) \over a(t)}~~\ .
\label{veldis}
\end{equation}
The average quantum state occupation number of the cold axions is
therefore
\begin{equation}
{\cal N} \sim {(2 \pi)^3~ n(t) \over {4 \pi \over 3} (m \delta v(t))^3}
\sim 10^{61} \left({f \over 10^{12}~{\rm GeV}}\right)^{8 \over 3}~~\ .
\label{phspden}
\end{equation}
${\cal N}$ is time-independent, in agreement with Liouville's theorem.
Considering that the axions are highly degenerate, it is natural to ask 
whether they form a Bose-Einstein condensate \cite{CABEC,therm}. We 
discuss the process of Bose-Einstein condensation and its implications 
in the next section.

The thermalization and Bose-Einstein condensation of cold dark matter
axions is also discussed in refs. \cite{Davidson,Yamaguchi, Jaeckel,Guth}
with conclusions that do not necessarily coincide with ours in all respects.

\section{Bose-Einstein condensation}

Bose-Einstein condensation occurs in a fluid made up of a huge 
number of particles if four conditions are satisfied: 1) the 
particles are identical bosons, 2) their number is conserved,
3) they are highly degenerate, i.e. ${\cal N}$ is much larger 
than one, and 4) they are in thermal equilibrium.  Axion number 
is effectively conserved because all axion number changing processes, 
such as axion decay to two photons, occur on time scales vastly longer 
than the age of the universe.  So the axions produced by vacuum 
realignment clearly satisfy the first three conditions.  The fourth
condition is not obviously satisfied since the axion is very weakly 
coupled.  In contrast, for Bose-Einstein condensation in atoms, the 
fourth condition is readily satisfied whereas the third is hard to 
achieve.  The fourth condition is a matter of time scales.  Consider 
a fluid that satisfies the first three conditions and has a finite, 
albeit perhaps very long, thermal relaxation time scale $\tau$.  Then, 
on time scales short compared to $\tau$ and length scales large compared 
to a certain Jeans' length (see below) the fluid behaves like cold dark 
matter (CDM), but on time scales large compared to $\tau$, the fluid 
behaves differently from CDM.

Indeed, on time scales short compared to $\tau$, the fluid behaves 
as a classical scalar field since it is highly degenerate.  In the 
non-relativistic limit, appropriate for axions, a classical scalar 
field is mapped onto a wavefunction $\psi$ by
\begin{equation}
\phi(\vec{r},t) = \sqrt{2} Re[e^{- i m t} \psi(\vec{r},t)]~~\ .
\label{wvfn}
\end{equation}
The field equation for $\phi(x)$ implies the Schr\"odinger-Gross-
Pitaevskii equation for $\psi$
\begin{equation}
i \partial_t \psi = - {1 \over 2m} \nabla^2 \psi + V(\vec{r},t) \psi
\label{SGP}
\end{equation}
where the potential energy is determined by the fluid itself:
\begin{equation}
V(\vec{r},t) = m \Phi(\vec{r},t) - {\lambda \over 8 m^2}|\psi(\vec{r},t)|^2~~\ .
\label{pot}
\end{equation}
The first term is due to the fluid's gravitational self-interactions.  
The gravitational potential $\Phi(\vec{r},t)$ solves the Poisson equation:
\begin{equation}
\nabla^2 \Phi = 4 \pi G m n~~\ ,
\label{Poisson}
\end{equation}
where $n=|\psi|^2$.  The fluid described by $\psi$ has density $n$ and 
velocity $\vec{v} = {1 \over m} \vec{\nabla} \arg(\psi)$.  Eq.~(\ref{SGP}) 
implies that $n$ and $\vec{v}$ satisfy the continuity equation and the 
Euler-like equation
\begin{equation}
\partial_t \vec{v} + (\vec{v}\cdot\vec{\nabla})\vec{v} = 
- {1 \over m} \vec{\nabla} V - \vec{\nabla} q
\label{Euler}
\end{equation}
where 
\begin{equation}
q = - {1 \over 2 m^2} {\nabla^2 \sqrt{n} \over \sqrt{n}}~~\ .
\label{q}
\end{equation}
$q$ is commonly referred to as `quantum pressure'.  The $\vec{\nabla} q$
term in Eq.~(\ref{Euler}) is a consequence of the Heisenberg uncertainty 
principle and accounts, for example, for the intrinsic tendency of a 
wavepacket to spread.  It implies a Jeans length \cite{Maxim}
\begin{equation}
\ell_J = (16 \pi G \rho m^2)^{-1 \over 4} = 1.01 \cdot 10^{14}~{\rm cm} 
\left({10^{-5}~{\rm eV} \over m}\right)^{1 \over 2}
\left({10^{-29}~{\rm gr/cm}^3 \over \rho}\right)^{1 \over 4}
\label{Jeans}
\end{equation}  
where $\rho = nm$ is the energy density.  On distance scales large 
compared to $\ell_J$, quantum pressure is negligible. CDM satisfies 
the continuity equation, the Poisson equation, and  Eq.~(\ref{Euler}) 
without the quantum pressure term.  So, on distance scales large 
compared to $\ell_J$ and time scales short compared to $\tau$, a 
degenerate non-relativistic fluid of bosons satisfies the same 
equations as CDM and hence behaves as CDM.  The wavefunction
describing density perturbations in the linear regime is given 
in ref. \cite{lin}.

On time scales large compared to $\tau$, the fluid of degenerate 
bosons does not behave like CDM since it thermalizes and forms a 
BEC.  Most of the particles go to the lowest energy state available 
to them through their thermalizing interactions.  This behavior is 
not described by classical field theory and is different from that 
of CDM.  When thermalizing, classical fields suffer from a ultraviolet 
catastrophe because the state of highest entropy is one in which each 
field mode has average energy $k_B T$ where $T$ is temperature.  In 
contrast, for the quantum field, the average energy of each mode is 
given by the Bose-Einstein distribution, and the ultraviolet catastrophe 
is removed.  To see whether Bose-Einstein condensation is relevant to
axions one must estimate the relaxation rate $\Gamma \equiv {1 \over \tau}$ 
of the axion fluid.  We do this in the next section.

When the mass is of order $10^{-21}$ eV and smaller, the Jeans 
length is long enough to affect structure formation in an observable 
way \cite{VLALP}.  Because we are focussed on the properties of QCD 
axions, we do not consider this interesting possibility here.

\section{Thermalization rate}

It is convenient to introduce a cubic box of size $L$ with periodic 
boundary conditions.  In the non-relativistic limit, the Hamiltonian 
for the axion fluid in such a box has the form 
\begin{equation}
H = \sum_j \omega_j a_j^\dagger a_j +
\sum_{j,k,l,m} {1 \over 4} \Lambda_{jk}^{lm}
a_j^\dagger a_k^\dagger a_l a_m
\label{Hamil}   
\end{equation}
with the oscillator label $j$ being the allowed particle momenta in 
the box $\vec{p} = {2 \pi \over L}(n_x,n_y,n_z)$, with $n_x$, $n_y$ 
and $n_z$ integers, and the $\Lambda_{jk}^{lm}$ given by \cite{therm}
\begin{equation}
\Lambda_{\vec{p}_1,\vec{p}_2}^{\vec{p}_3,\vec{p}_4} =
\Lambda_{s~\vec{p}_1,\vec{p}_2}^{~~\vec{p}_3,\vec{p}_4}~~+~~
\Lambda_{g~\vec{p}_1,\vec{p}_2}^{~~\vec{p}_3,\vec{p}_4}   
\label{Lam}
\end{equation}
where the first term
\begin{equation}
\Lambda_{s~\vec{p}_1,\vec{p}_2}^{~~\vec{p}_3,\vec{p}_4}
= - {\lambda \over 4 m^2 L^3}~
\delta_{\vec{p}_1 + \vec{p}_2, \vec{p}_3 + \vec{p}_4}
\label{selfc}
\end{equation}
is due to the $\lambda \phi^4$ self-interactions, and the 
second term
\begin{equation}
\Lambda_{g~\vec{p}_1,\vec{p}_2}^{~~\vec{p}_3,\vec{p}_4}
= - {4 \pi G m^2 \over L^3}
\delta_{\vec{p}_1 + \vec{p}_2, \vec{p}_3 + \vec{p}_4}~
\left({1 \over |\vec{p}_1 - \vec{p}_3|^2}
+ {1 \over |\vec{p}_1 - \vec{p}_4|^2}\right)
\label{gravc}
\end{equation}
is due to the gravitational self-interactions.  

In the particle kinetic regime, defined by the condition that the 
relaxation rate $\Gamma \equiv {1 \over \tau}$ is small compared 
to the energy dispersion $\delta \omega$ of the oscillators, the 
Hamiltonian of Eq.~(\ref{Hamil}) implies the evolution equation 
\begin{equation}
\dot{\cal N}_l = \sum_{k,i,j= 1} {1 \over 2} |\Lambda_{ij}^{kl}|^2
\left[{\cal N}_i {\cal N}_j ({\cal N}_l + 1)({\cal N}_k + 1)   
- {\cal N}_l {\cal N}_k ({\cal N}_i + 1)({\cal N}_j + 1)\right]
2 \pi \delta(\omega_i + \omega_j - \omega_k - \omega_l)
\label{Bolq}
\end{equation}
for the quantum state occupation number operators
${\cal N}_l(t) \equiv a_l^\dagger(t)a_l(t)$.  The thermalization 
rate in the particle kinetic regime, is obtained by carrying out 
the sums in Eq.~(\ref{Bolq}) and estimating the time scale $\tau$
over which the ${\cal N}_j$ change completely.  This yields
\cite{CABEC,therm}
\begin{equation}
\Gamma \sim n~\sigma~\delta v~{\cal N}
\label{pkg} 
\end{equation}
where $\sigma$ is the scattering cross-section associated with the 
interaction, and ${\cal N}$ is the average state occupation number 
of those states that are highly occupied.  The cross-section for 
scattering by $\lambda \phi^4$ self-interactions is 
$\sigma_\lambda = {\lambda^2 \over 64 \pi m^2}$.  For gravitational 
self-interactions, one must take the cross-section for large angle 
scattering only, $\sigma_g \sim {4 G^2 m^2 \over (\delta v)^4}$, 
since forward scattering does not change the momentum distribution.  

However, the axion fluid does not thermalize in the particle kinetic 
regime.  It thermalizes in the opposite ``condensed regime" defined 
by $\Gamma >> \delta \omega$.  In the condensed regime, the relaxation 
rate due to $\lambda \phi^4$ self-interactions is \cite{CABEC,therm}
\begin{equation}
\Gamma_\lambda \sim {n \lambda \over 4 m^2} 
\label{gl}
\end{equation}
and that due to gravitational self-interactions is 
\begin{equation}
\Gamma_g \sim 4 \pi G n m^2 \ell^2
\label{gg}
\end{equation}
where $\ell = {1 \over m \delta v}$ is, as before, the correlation 
length of the particles.  One can show that the expressions for the 
relaxation rates in the condensed regime agree with those in the 
particle kinetic regime at the boundary $\delta \omega = \Gamma$.

We apply Eqs.~(\ref{gl}) and (\ref{gg}) to the fluid of cold 
dark matter axions described at the end of Section 1.  One finds  
that $\Gamma_\lambda(t)$ becomes of order the Hubble rate, and 
therefore the axions briefly thermalize as a result of their 
$\lambda \phi^4$ interactions, immediately after they are produced 
during the QCD phase transition.  This brief period of thermalization 
has no known implications for observation.  However, the axion fluid 
thermalizes again due to its gravitational self-interactions when the 
photon temperature is approximately \cite{CABEC,therm}
\begin{equation}
T_{\rm BEC} \sim 500~{\rm eV} 
\left({f \over 10^{12}~{\rm GeV}}\right)^{1 \over 2}~~\ .
\label{Tbec}
\end{equation}
The axion fluid forms a BEC then.  After BEC formation, the correlation 
length $\ell$ increases till it is of order the horizon and thermalization 
occurs on ever shorter time scales relative to the age of the universe.

\section{Observational consequences}

As was emphasized in Section 3, the axion fluid behaves differently 
from CDM when it thermalizes.  Indeed when all four conditions for 
Bose-Einstein condensation the axions almost all go to their lowest 
energy available state. CDM does not do that.  One can readily show 
that in first order of perturbation theory and within the horizon 
the axion fluid does not rethermalize and hence behaves like CDM.  
This is important because the cosmic microwave background observations 
provide very strong constraints in this arena and the constraints are 
consistent with CDM.  In second order of perturbation theory and higher, 
axions generally behave differently from CDM.  The rethermalization of 
the axion BEC is sufficiently fast that axions that are about to fall 
into a galactic gravitational potential well go to their lowest energy 
state consistent with the total angular momentum they acquired from 
nearby protogalaxies through tidal torquing \cite{therm}.  That state 
is a state of net overall rotation.  In contrast, CDM falls into 
galactic gravitational potential wells with an irrotational velocity 
field.  The inner caustics are different in the two cases.  In the 
case of net overall rotation, the inner caustics are rings \cite{crdm} 
whose cross-section is a section of the elliptic umbilic $D_{-4}$ 
catastrophe \cite{sing}, called caustic rings for short.  If the 
velocity field of the infalling particles is irrotational, the inner 
caustics have a `tent-like' structure which is described in detail 
in ref.~{\cite{inner} and which is quite distinct from caustic rings.  
There is observational evidence for caustic rings \cite{MWha}.  It was 
shown \cite{case} that the assumption that the dark matter is axions 
explains not only the existence of caustic rings but also their detailed 
properties, in particular the pattern of caustic ring radii and their 
overall size.  Furthermore, it was shown \cite{angmom} that axion BEC 
solves the galactic angular momentum problem, the tendency of CDM to 
produce halos that are too concentrated at the center compared to 
observations.  

In a recent paper \cite{Newberg}, J. Dumas et al. compare the 
predictions of the caustic ring model with the rotation curve 
of the Milky Way and the observations of the Sagittarius sattelite's 
tidal disruption.

\section{Acknowledgments}

We would like to thank Joerg Jaeckel, Alan Guth, Mark Hertzberg and 
Chanda Prescod-Weinstein for stimulating discussions.  This work was
supported in part by the US Department of Energy under grant 
DE-FG02-97ER41209.

\section{Bibliography}

\begin{footnotesize}

\end{footnotesize}


\end{document}